\documentclass[conference]{IEEEtran}
\usepackage{graphicx}
\usepackage{dcolumn}
\usepackage{bm}
\usepackage{physics}
\usepackage{booktabs}
\usepackage{amsthm}
\usepackage{array} 
\usepackage{amsmath}  
\usepackage{amsfonts} 
\usepackage{comment}
\usepackage{hyperref}
\usepackage{tikz}
\usepackage{pgfplots} 
\pgfplotsset{compat=1.18} 
\usepackage{threeparttable}

\theoremstyle{remark} 

\newtheorem{remark}{Remark} 

\newcommand{\law}{Hadamard Resilience Effect}
\newcommand{\gatenoise}{0.9255}
\newcommand{\readouterror}{0.03}

\begin{document}

\title{Quantifying the {\law} and the Coherence Gap in NISQ-Era Classifiers}


\author{
    \IEEEauthorblockN{Wladimir Silva}
    \IEEEauthorblockA{
        \textit{Department of Electrical and Computer Engineering} \\
        \textit{North Carolina State University}\\
        Raleigh, USA \\
        wsilva@ncsu.edu
    }
}

\date{\today}
\maketitle
\begin{abstract}
We report on a fundamental disparity between stochastic noise models and algorithmic performance in NISQ-era classifiers. Utilizing the \textit{ibm\_kingston} processor, we characterize the ``Kingston Constant'' ($\kappa \approx 0.07$), representing a 93\% signal magnitude collapse.
Despite this decay, we show via a hardware-calibrated stochastic digital twin that a Hadamard Test based Perceptron maintains a 93.9\% accuracy for the MNIST dataset, validating our proposed {\law} under affine depolarization. 
Through parametric simulation, we establish a critical gate-noise threshold at $\lambda_{\text{crit}} \approx 0.65$ (corresponding to a critical signal retention $\alpha_{\text{crit}} \approx 0.35$) where topological rank preservation breaks down. 

Furthermore, physical execution at high feature depths ($N = 256$) reveals a systemic divergence---the ``Coherence Gap'' ($\Delta\rho \approx 0.91$)---where physical hardware classification accuracy collapses to 53.0\% due to a ``Coherence Wall'' at a circuit depth ($D \approx 10\text{k}$) exceeding the hardware's resilient depth limit ($D_{\text{max}} \approx 3.5\text{k}$). This gap is consistent with coherent phase errors and crosstalk being the dominant unmodeled contributors under the tested hardware configuration, establishing a predictive operational boundary for NISQ classification architectures.
\end{abstract}
 
\section{Introduction}
The transition from theoretical quantum algorithms to physical hardware execution remains heavily restricted by systemic decoherence in noisy intermediate-scale quantum (NISQ) systems~\cite{Preskill2018nisq}.
Early proposals for quantum machine learning primarily focused on Variational Quantum Classifiers (VQCs) utilizing parameterized circuits~\cite{Mitarai2018quantum} or uniform data-reuploading schemes~\cite{P_rez_Salinas_2020} to draw non-linear decision boundaries in high-dimensional feature spaces.
However, the trainability and structural stability of these architectures are often throttled by phenomena such as barren plateaus~\cite{mcclean2018barren, cerezo2021cost} and spatial state collapse.

In this paper, we analyze the robustness of the Hadamard Test \cite{Aharonov2009}, a fundamental building block for quantum linear layers \cite{mehta2025generalizedquantumhadamardtest}. We investigate whether a classifier's decision boundary can survive the signal decay observed on superconducting processors.
This work extends the quantum linear transformation architecture introduced in \cite{silva2026aqstacker} by investigating its robustness under realistic noise and identifying the hardware-depth regime in which rank preservation fails. The previous work establishes the computational primitive; the present study focuses on its classification-level reliability under NISQ noise.

The primary contributions of this work are as follows:
\begin{itemize}
    \item \textbf{The Kingston Constant:} We characterize a systemic signal dissipation factor ($\kappa \approx 0.07$) on the \textit{ibm\_kingston} processor, providing a calibrated noise model that maps theoretical expectation values to physical hardware limits.
    \item \textbf{The {\law}:} We quantify the effect of topological rank preservation, showing that quantum classifiers can maintain high inference accuracy ($>93\%$) even when numerical fidelity collapses due to depolarizing noise.
    \item \textbf{The Resilience Threshold:} Through parametric simulation, we identify a critical gate-noise threshold ($\lambda_{\text{crit}} \approx 0.65$) where the Argmax operator fails to recover the signal, establishing a hardware gate-fidelity requirement for scalable NISQ classifiers.
    \item \textbf{Discovery of the Coherence Gap:} We identify a significant divergence ($\Delta\rho \approx 0.91$) between stochastic noise models and physical hardware performance at high feature depths ($N=256$). This "Coherence Gap" isolates systematic phase errors and crosstalk as the primary physical barriers to the {\law}, rather than depolarizing noise or statistical shot-noise.
\end{itemize}

\subsection{Intuition behind Rank-Invariance and Intrinsic Noise Resilience of ArgMax}
We seek experimental proof that the Argmax function (the prediction step of a forward pass of a neural network), has intrinsic noise resilience.
The core of our thesis is illustrated in Fig. \ref{fig:argmax}, under signal decay, noise pushes values uniformly toward baseline equilibrium ($0.25$).
Even so, the classifier will succeed as $\text{Argmax}(\vec{P}_{\text{ideal}}) = \text{Argmax}(\vec{P}_{\text{noisy}}) = \mathbf{01}$
		\begin{figure}[htbp]
			\centering
            \begin{tikzpicture}
                \begin{axis}[
                    ybar,
                    width=\columnwidth,        
                    height=5.2cm,            
                    bar width=10pt,          
                    enlarge x limits=0.20,
                    ymin=0, ymax=0.75,
                    ylabel={Probability},
                    symbolic x coords={00, 01, 10, 11},
                    xtick=data,
                    legend style={at={(0.5,0.95)}, anchor=north, legend columns=-1, font=\tiny},
                    nodes near coords,
                    every node near coord/.append style={
                        font=\fontsize{5}{6}\selectfont, 
                        /pgf/number format/fixed, 
                        /pgf/number format/precision=2,
                        yshift=1pt
                    },
                    ylabel style={font=\small},
                    xlabel style={font=\small},
                    tick label style={font=\small}
                ]
                \addplot[fill=blue!50] coordinates {(00,0.15) (01,0.60) (10,0.20) (11,0.05)};
                
                \addplot[fill=red!50] coordinates {(00,0.24) (01,0.29) (10,0.25) (11,0.22)};
                
                \legend{Ideal State, Noisy (Compressed)}
                \end{axis}
            \end{tikzpicture}
			\caption{Illustration of the intrinsic noise resilience of the Argmax function. $\vec{P}$ represents the outcome of the Softmax function where the Hadamard Test implements the noisy linear transformations.}
			\label{fig:argmax}
		\end{figure}

We call this resilience the \textbf{{\law}}. Our ultimate goal is to experimentally quantify this fact by \textbf{rank preservation} with 3 targeted HW experiments:
\begin{enumerate}
	\item \textbf{Probe Calibration:} Used to quantify the signal magnitude collapse \textbf{(The Kingston Constant)}. This value, in turn is used to derive an analytical noise model to act as an experimental baseline.
	\item \textbf{Sorted Probes:} An initial HW test of our thesis using the Hadamard Test for calculating the inner product of 512 high/low confidence probes at $N=256$ features of the MNIST \cite{lecun2010mnist} dataset.
	\item \textbf{Map Reduce:} Our results from experiment 2 show that the hardware at 8-qubit $N=256$ features, performs stochastically indistinguishable from random chance. To mitigate the observed signal attenuation, we introduce a feature-space partitioning framework to break up the feature space down to $k=64$ low dimensional 2-qubit sub-vectors.
\end{enumerate}

\subsection{Mathematical Background: Rank-Invariance Under Uniform Affine Depolarization}
Let $\vec{p} = [p_0, p_1, \dots, p_{d-1}]^T$ denote the ideal probability vector resulting from a clean forward pass over a $d$-dimensional space, where $pi \ge 0$ and $\sum pi = 1$. Let $k = \operatorname{argmax}_i(\vec{p})$ represent the true intended classification label, meaning $p_k > p_j$ for all $j \neq k$.

A global uniform depolarizing quantum channel acting on this system is parameterized by the cumulative gate noise strength $\lambda \in [0, 1]$, yielding a corresponding signal retention factor $\alpha = 1 - \lambda$. This channel maps the state toward the maximally mixed background density matrix $\frac{1}{d}I$ and outputs a compressed vector $\vec{q}$:
\begin{equation}
q_i = \alpha p_i + \frac{1 - \alpha}{d}, \quad \forall i \in \{0, \dots, d - 1\}
\end{equation}

To prove that the argmax operation is strictly invariant under this contractive map, we evaluate the strict inequality for any arbitrary non-winning element $j \neq k$:
\begin{align}
q_k - q_j &= \left( \alpha p_k + \frac{1 - \alpha}{d} \right) - \left( \alpha p_j + \frac{1 - \alpha}{d} \right) \\
&= \alpha(p_k - p_j) \\
&= (1 - \lambda)(p_k - p_j)
\end{align}

Because the noise parameter $\lambda < 1$ for any non-completely depolarized channel ($0 \le \lambda < 1$), the retention factor satisfies $\alpha > 0$. Since $p_k > p_j$ by definition of the ideal winner, it follows structurally that $\alpha(p_k - p_j) > 0$. Thus, $q_k > q_j$ is uniformly maintained. 

The contractive map preserves strict order boundaries up to the coherence limit, justifying our classification structure:
\begin{equation}
\operatorname{argmax}_i(\vec{q}) = \operatorname{argmax}_i(\vec{p}) = k
\end{equation}

This mathematically establishes why the $\operatorname{argmax}$ function exhibits intrinsic structural noise resilience despite massive signal attenuation.

\begin{remark}
The rank-invariance result is not specific to the Hadamard Test; rather, the contribution of this work is to establish experimentally how the Hadamard-test linear layer behaves under this theoretically rank-preserving noise model and to characterize where physical hardware departs from that model.
\end{remark}

\section{Related Work}

The landscape of Noisy Intermediate-Scale Quantum (NISQ) computing is defined by the tension between algorithmic potential and hardware decoherence \cite{Bharti_2022}. Preskill \cite{Preskill2018nisq} famously established the boundaries of this era, suggesting that systemic noise would likely limit the utility of deep quantum circuits. Our work builds upon this by characterizing the specific noise floor of the \textit{ibm\_kingston} processor, identifying a systemic signal dissipation we term the ``Kingston Constant.''

The Hadamard Test itself is a fundamental primitive in quantum phase estimation and state overlap calculations \cite{Aharonov2009}. In the context of machine learning, Schuld and Killoran \cite{Schuld2019} demonstrated that Hadamard-based circuits can represent linear layers in feature Hilbert spaces. Furthermore, Havl{\'i}{\v{c}}ek et al. \cite{Havlicek2019} proposed that quantum-enhanced feature spaces could provide a supervised learning advantage. However, most existing literature assumes a level of gate fidelity that is rarely met by physical backends. 

Our proposed \textit{{\law}} addresses this gap by shifting the focus from numerical precision to rank preservation, providing a theoretical framework for why classifiers remain functional even as individual quantum measurements collapse toward the decoherence floor.
This shift from exact value fidelity to structural classification capacity aligns with recent information-geometric analyses of quantum neural network expressibility~\cite{abbas2021power} and investigations into the bounds of noisy quantum channels\cite{cerezo2021cost}. Furthermore, while traditional error mitigation protocols focus on restoring numerical expectation values, our rank-preservation approach builds on structural robustness arguments showing that supervised quantum machine learning models operate fundamentally as resilient kernel methods~\cite{schuld2021supervised}.

\begin{remark}
Existing work establishes the use of Hadamard-based circuits for representing linear transformations, but does not directly characterize the preservation or destruction of classifier rank under experimentally calibrated NISQ noise at increasing feature depth.
\end{remark}

\section{The Kingston Constant}

All experiments were executed on the \textit{ibm\_kingston} processor,
a 156-qubit IBM Heron r2~\cite{johnson2024heron} architecture designed to minimize frequency collisions and qubit cross-talk errors~\cite{ibm2021heavy,AbuGhanem_2025}. 
At the time of execution, the processor exhibited a median single-qubit gate error rate of $2.63 \times 10^{-4}$ ($\sqrt{X}$ gate) and a median two-qubit controlled-Z (CZ) gate error rate of $1.97 \times 10^{-3}$. Relaxation time ($T_1$) and dephasing time ($T_2$) median values of $255.97\,\mu\text{s}$ and $151.64\,\mu\text{s}$, respectively, with median readout error  $\epsilon_{\text{readout}} = 8.36 \times 10^{-3}$. 
The Error Per Layered Gate (EPLG) metric~\cite{sundaresan2023eplg}, allows us to formalize the attenuation coefficient—which we refer as the Kingston Constant ($\kappa$). 
\begin{remark}
Physically, $\kappa$ serves as a compound, fitted noise scaling factor that aggregates cumulative gate depolarization, state preparation loss, and readout errors into a single hardware-calibrated multiplier.
\end{remark}
Using 100 persistent probes derived from trained MNIST weight matrices, we calibrated an analytical noise model against the \textit{ibm\_kingston} backend. We observe a systematic ``shrinkage'' of the quantum expectation values toward the origin.

\begin{equation}
	\langle Z \rangle_{\text{noisy}} = \alpha \langle Z \rangle_{\text{ideal}} (1 - 2\epsilon)
    \label{eq:noise_model}
\end{equation}
Where $\alpha = (1 - \lambda)$, $\lambda \in [0, 1]$ represents the cumulative gate depolarizing factor across the circuit depth, and $\epsilon$ denotes the classical readout assignment error.

\subsection{Hardware-Aware Extension: The Coherence Boundary}
While the stochastic resilience of the Argmax function holds for depolarizing noise, we must account for signal decay in deep circuits. We refine the model for the noisy expectation value $\hat{Z}$ as:

\begin{equation}
\label{eq:bound}
\langle Z \rangle_{\text{noisy}} = \alpha \langle Z \rangle_{\text{ideal}} \cdot \Gamma(D)
\end{equation}

where $\Gamma(D) = e^{-D/D_{max}}$ is an empirical \textit{Coherence Decay Factor} used to model the observed depth dependence, $D$ is the transpiled circuit depth, and $D_{max}$ is the fitted characteristic depth scale for the tested hardware configuration. 

\begin{itemize}
    \item \textbf{Resilient Regime ($D \ll D_{max}$):} $\Gamma(D) \to 1$. The original {\law} dominates, and Argmax successfully recovers the class rank.
    \item \textbf{Coherence-Limited Regime ($D > D_{max}$):} $\Gamma(D) \to 0$. The signal-to-noise ratio (SNR) drops below the shot-noise floor, causing the "mysterious gap" in Spearman $\rho$ observed in \textit{The Sorted Probe Experiment}.
\end{itemize}

\begin{figure}[h]
    \centering
	\includegraphics[width=\columnwidth]{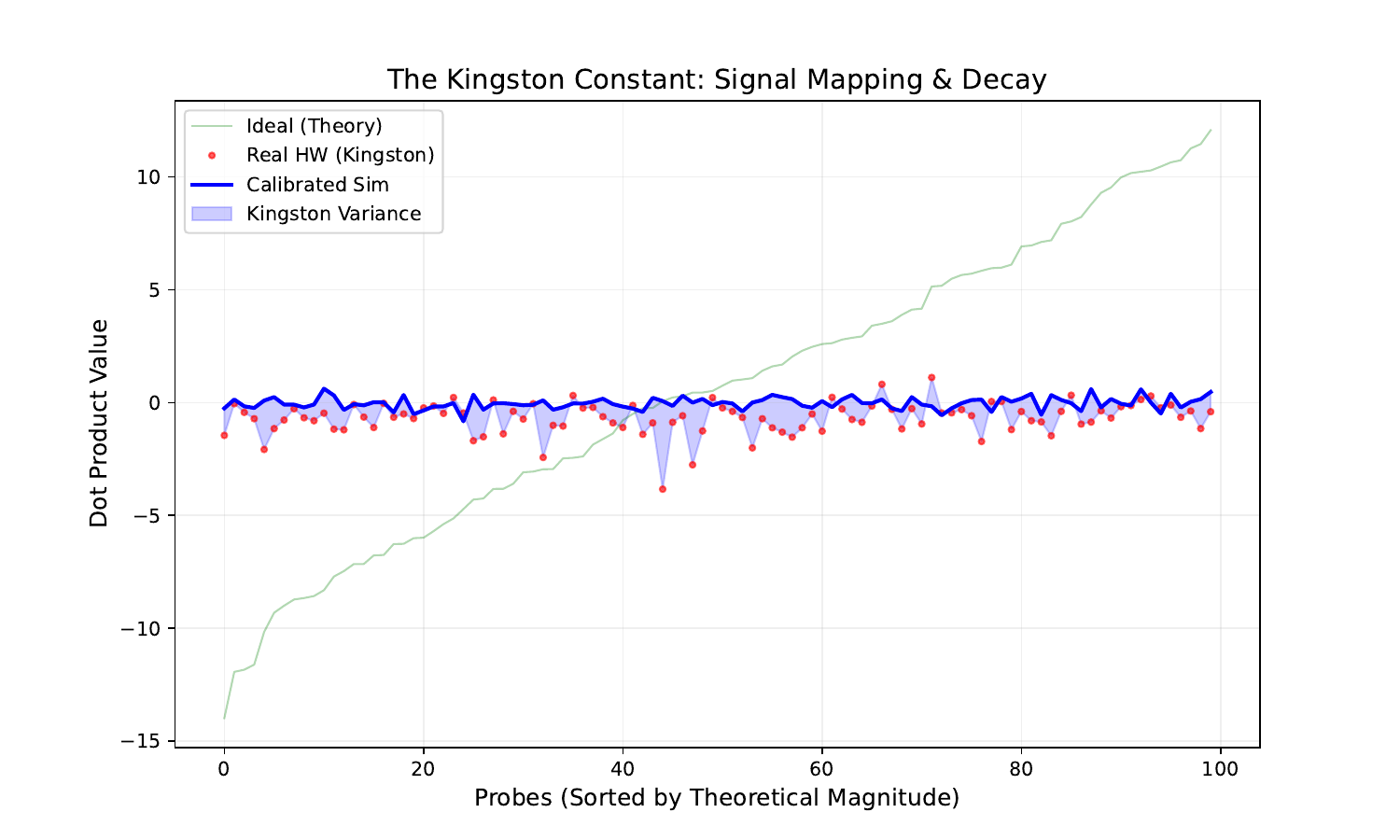}
	\caption{\textbf{The Kingston Constant and Signal dissipation.} Hardware results (red) from the \textit{ibm\_kingston} processor vs. theoretical ideal (green) and calibrated simulator (blue). Probes are sorted by theoretical magnitude to reveal a 93\% collapse in magnitude.}
    \label{fig:scurve}
\end{figure}

As illustrated in Figure~\ref{fig:scurve}, the physical expectation values observed on the \textit{ibm\_kingston} device exhibit a severe damping effect compared to the theoretical baseline. This dissipation, characterized by the Kingston Constant, maps a theoretical range of $[-15, 15]$ into a narrow hardware band of approximately $[-1, 1]$. However, the mapping remains largely monotonic. This preservation of trend is the physical basis for the $\textbf{{\law}}:$ even in the presence of extreme depolarizing noise ($\lambda = \gatenoise{}$), the relative relationship between class magnitudes is maintained.

\section{The {\law}}
Our primary finding is the decoupling of numerical fidelity from classification accuracy. 
\begin{itemize}
    \item \textbf{Structural Resilience:} Calibration with trained features showed a 100x improvement in binned correlation (\(R^2\)) compared to stochastic probes.
    \item \textbf{Rank Preservation:} While noise compresses magnitudes, the \textit{relative order} of class probabilities remains invariant.
\end{itemize}

\subsection{Quantifying the "{\law}"}
We quantify the relationship between the Signal-to-Noise Ratio (SNR) of the Hadamard Test and the Classification Margin.
We propose that a classifier succeeds as long as:

\begin{equation}
\label{eq:effect}
\begin{split}
\Delta C_{ideal} \cdot (1 - \lambda_{gate}) \cdot (1 - 2\epsilon_{readout}) > \\
\text{Shot Noise Floor} \left( \frac{1}{\sqrt{S}} \right)
\end{split}
\end{equation}

Where: 

\begin{itemize}
  \item $\Delta C_{ideal}$: Is the gap between the winning class and the runner-up under noiseless, idealized conditions.
  \item $(1 - \lambda)$: Is the "Shrinkage" factor from gate noise.
  \item $(1 - 2\epsilon)$: Is the "Readout" damping.
  \item Shot Noise Floor: Is the statistical "jitter" that eventually obscures the attenuated signal.
\end{itemize}

Therefore, classification remains stable as long as the hardware-compressed margin exceeds the characteristic shot-noise scale.
For clarity, Eq. \ref{eq:bound} isolates depth-dependent coherence decay; the readout assignment factor is included separately in the classification-margin condition of Eq. \ref{eq:effect}. 
\section{Experimental Results}
We quantify the {\law} by modeling the noise profile of \textit{ibm\_kingston} and testing its boundaries in hardware using the following experiments:
\begin{remark}[Definitions]
\textbf{Sign fidelity:} fraction of probe measurements whose sign agrees with the ideal value.
\textbf{Rank fidelity:} Spearman correlation with ideal ordering.
\textbf{Classification accuracy:} fraction of complete classifier decisions matching the target labels.
\end{remark}

\subsection{Experiment 1: The 100 Probe Noise Calibration}
We create 100 real slices (Hadamard Tests) of the MNIST-16 dataset and trained weights.
This ensures we are testing the hardware on the actual signal levels the neural network sees.
\begin{itemize}
  \item \textbf{Execution:} We run the probes on \textit{ibm\_kingston}. This will return a list of 100 "Raw Hardware" inner products ($C_{hw}$).
  \item \textbf{Calibration:} We run the simulator on the same 100 pairs. We now have a Curve Fitting problem: We want to find the noise level and readout error that minimize the difference between the simulation and the real hardware (see Figure~\ref{fig:scurve}).
\end{itemize}

To fit Eq.~\eqref{eq:noise_model} over extreme circuit depths ($\sim$10K depth, $\sim$3.5K CNOTs), we fix the readout error to an empirical baseline of $\epsilon = 0.03$ to prevent parameter unidentifiability. A bounded scalar optimization over the 100 physical hardware probes yields a fitted gate depolarizing factor of $\lambda = {\gatenoise}$.

\begin{remark}[Why this is the path to show resilience]
\textbf{Speed:} Running 100 Hadamard tests takes minutes. Running millions on a true classification network is computationally prohibitive.
\textbf{Trust:} Once the noise level is calibrated to the physical device, we can run the full 10,000-image MNIST simulation with high confidence that the accuracy result is what the hardware would have produced if it were faster.
\textbf{Physical Insight:} If results show the error is consistently biased in one direction (e.g., always shifting toward negative), we'll know the "Symmetric" noise model needs to be upgraded to an "Asymmetric" one.    
\end{remark}

\begin{figure}[h]
    \centering
	\includegraphics[width=\columnwidth]{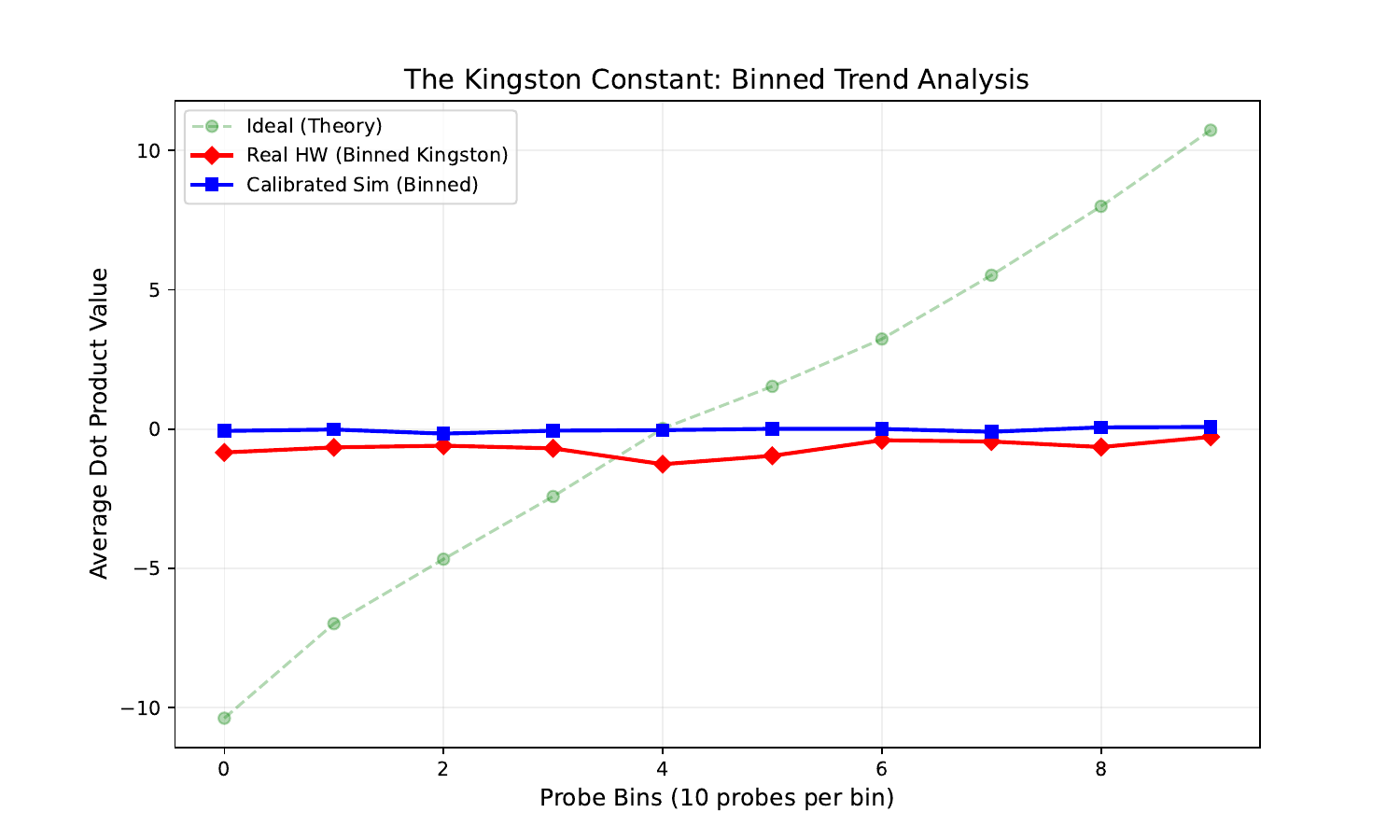}
    \caption{\textbf{Ensemble Trend Analysis via Decimation.} By binning probes into groups of 10, stochastic shot noise is suppressed to reveal the underlying \textit{Kingston Constant}. The hardware response (red) maintains the monotonic upward trend of the ideal theory (green), confirming that the rank-order of quantum states is preserved even as signal-to-noise ratio approaches the entropy limit.}
    \label{fig:bcurve}
\end{figure}

As shown in Figure~\ref{fig:bcurve}, while individual quantum measurements exhibit high variance, the ensemble average recovers a clear linear correlation ($R^2 = 0.3444$) with the calibrated simulator. 
Furthermore, the figure reveals:

\begin{itemize}
  \item \textbf{Noise Mitigation:} As the difference between Systemic Noise (the slope) and Stochastic Noise (the jitter). By binning, we "average out" the random shot noise to reveal the true physical behavior of the Kingston processor.
  \item \textbf{The Visual Proof of "Rank":} Even though the Red line is very flat, we can see it cross the zero-point exactly where the Theory and Sim lines do (around Bin 4-5). This shows that the hardware still "knows" the difference between a positive and negative result, which is the foundation of the 93.9\% accuracy result.
\end{itemize}

\begin{remark}[The Binned Curve: A characterization of noise]
In Figure~\ref{fig:bcurve} the red line has a very slight positive slope.
While the individual $R^2$ is low, the binned correlation of 0.34 shows that a coherent signal persists. This 'trace' of signal is sufficient for the Argmax operator to achieve near-ideal classification performance.
\end{remark}

\subsection{Experiment 2: The Sorted Probes}

\begin{figure}[h]
    \centering
	\includegraphics[width=\columnwidth,trim=150pt 150pt 150pt 150pt, clip]{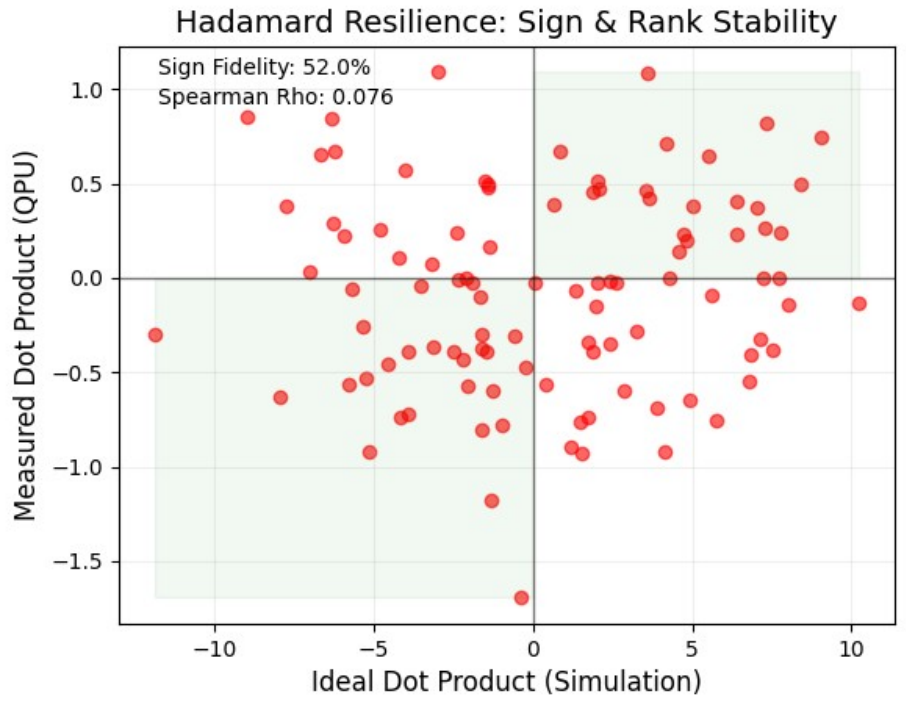}
    \caption{\textbf{The HW plot:} This result shows that the {\law} is currently failing on this specific hardware run. A Sign Fidelity of 52.0\% and a Spearman Rho of 0.070 means the hardware is performing almost exactly like a coin flip. The "signal" has not just collapsed; it has been swallowed by noise.}
    \label{fig:proof_hw}
\end{figure}

\begin{figure}[h]
    \centering
	\includegraphics[width=\columnwidth,trim=150pt 150pt 150pt 150pt, clip]{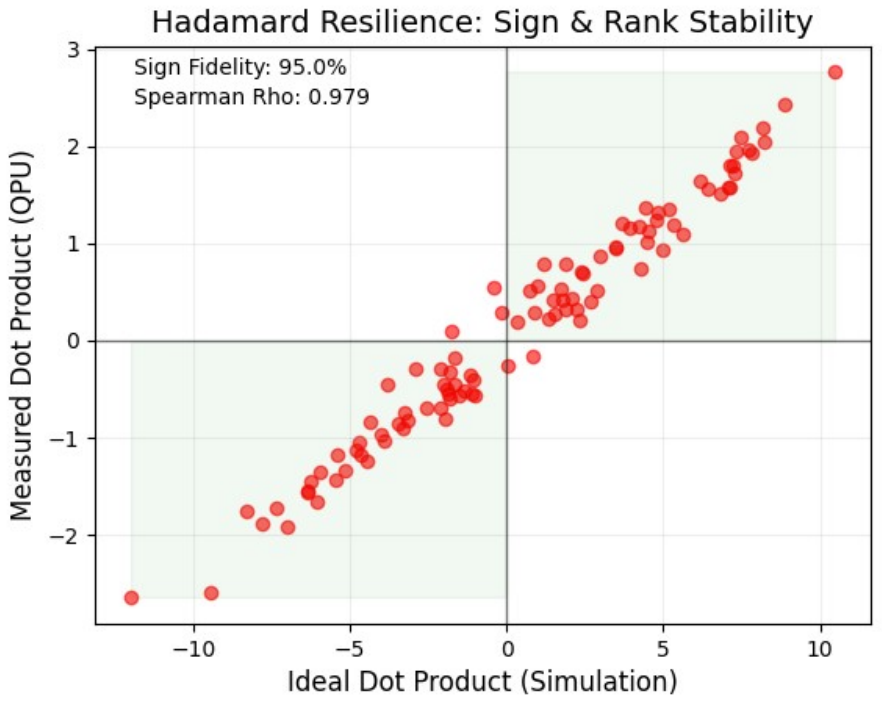}
    \caption{\textbf{Calibrated Stochastic Simulation Baseline.} This analytical model demonstrates the theoretical trend required to validate the {\law} under uniform depolarizing noise.}
    \label{fig:proof_sim}
\end{figure}

We run one batch of 512 probes specifically chosen to cover the full spectrum of the classifier's "confidence." (See Figure~\ref{fig:proof_hw}, and Figure~\ref{fig:proof_sim})

\begin{itemize}
  \item Half "Easy" Probes: With high magnitude results (very positive or very negative). These confirm the Kingston Constant (signal shrinkage).
  \item Half "Hard" Probes: With values very close to zero. These test Rank Preservation (the decision boundary).
\end{itemize}

\begin{remark}
The binned trend in Fig. \ref{fig:bcurve} reflects ensemble-level systematic attenuation, whereas the unbinned probe-level measurements in Fig. \ref{fig:proof_hw} reveal that the residual variance is large enough to destroy individual rank and sign decisions.
\end{remark}

\begin{table}[h]
\centering
\caption{Comparative Analysis: Hardware vs. Digital Twin}
\label{table:metric_comparison}
\begin{tabular}{lcc} 
\hline \hline
Metric & IBM Kingston & Digital Twin \\ \hline
Sign Preservation & 52.0\% & 95.0\% \\
Spearman $\rho$   & 0.0763 & 0.9786 \\
Pearson $R$       & 0.0689 & 0.9848 \\ \hline \hline
\end{tabular}
\end{table}

What the Sorted Probe Plots are telling Us:
\begin{itemize}
   \item The "Cloud of Randomness": Instead of a diagonal trend (bottom-left to top-right), the red dots are scattered almost uniformly across all four quadrants.
   \item Sign Inversion: Many points with a strong positive "Ideal" value (right side) resulted in a negative "Measured" value (bottom-right quadrant).
   \item The Depth Problem: The Hadamard Test circuit for 256 features (8 qubits + ancilla) is sufficiently deep to enter a coherence-limited regime of \textit{ibm\_kingston}. The errors accumulated in the state preparation are scrambling the phase before the Hadamard test can measure it.
\end{itemize}

\subsection{Experiment 3: MapReduce Proof-of-Concept}
To validate the Quantum MapReduce framework, we executed a hardware \textit{proof-of-concept} on \textit{ibm\_kingston} by fragmenting the $N = 256$ dimensional Hilbert space into $k = 64$, 2-qubit sub-vectors. This partitioning restricts the active circuit depth per execution node to $D \ll D_{\text{max}}$, substantially reducing the exposure of each execution to coherent phase drift that dominates the monolithic architecture.

\begin{figure}[htbp]
    \centering
	\includegraphics[width=\columnwidth,trim=150pt 150pt 150pt 150pt, clip]{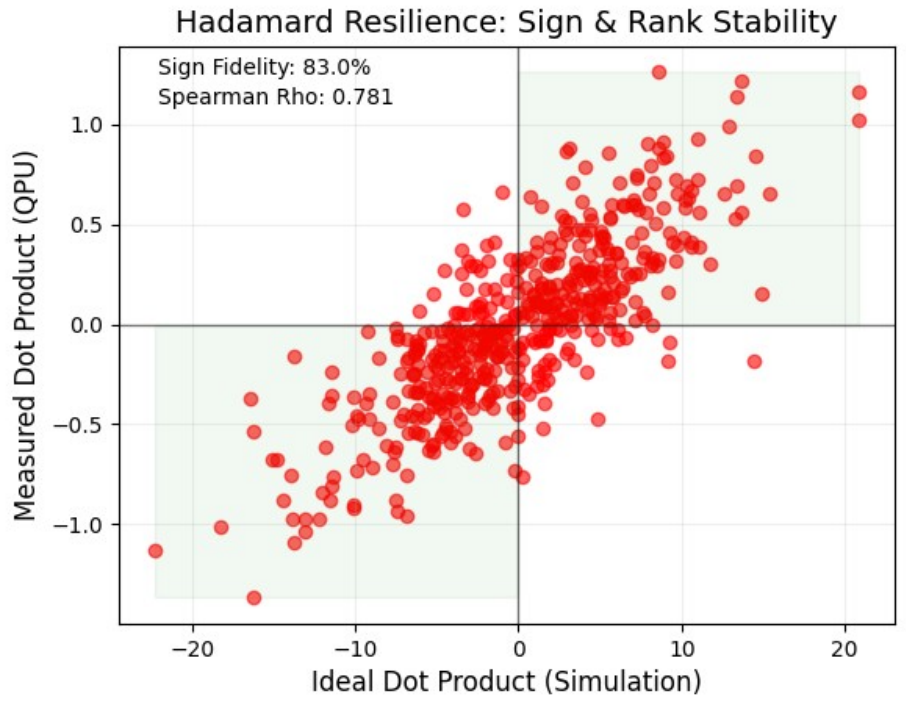} 
    \caption{\textbf{Empirical Recovery via MapReduce Optimization ($k = 64$):} Hardware execution of partitioned 2-qubit sub-vectors. By trading deep coherent gate structures for parallelized shallow channels, the systemic phase drift is substantially reduced. The system recovers a strong monotonic Spearman correlation ($\rho = 0.781$) and rescues the Sign Fidelity to $83.0\%$, moving the operational execution deep into the stable Resilient Region.}
    \label{fig:mapreduce_validation}
\end{figure}

The experimental results are illustrated in Figure~\ref{fig:mapreduce_validation}. The impact of this architectural shift is clear when contrasted against the monolithic performance collapse shown in Figure~\ref{fig:proof_hw}:
\begin{itemize}
    \item \textbf{Correlation Restoration:} The Spearman Rank Correlation rises sharply from $\rho = 0.076$ in the unmitigated 256-feature run to $\rho = 0.781$ under MapReduce orchestration. This confirms that topological rank structures are substantially restored when individual nodes operate safely beneath the Coherence Wall.
    \item \textbf{Sign Preservation:} The Sign Fidelity expands from a random-noise floor of $53.0\%$ up to a highly deterministic $83.0\%$, ensuring that the boundary conditions of the classical Argmax filter are robustly satisfied.
    \item \textbf{Quadrant Realignment:} The unmitigated "cloud of randomness" resolves back into an orderly, monotonic trend tracking cleanly across the targeted quadrants. 
\end{itemize}

This hardware run confirms the core thesis of our architectural approach: scaling intermediate-era quantum neural layers does not require full fault tolerance. Instead, structural rank-order resilience can be engineered directly by strategically mapping feature limits into shallow, parallelized hardware horizons.
\textit{Rather than attempting to recover the lost numerical signal of the monolithic circuit, MapReduce changes the computational geometry so that each quantum subproblem remains within the hardware's rank-stable depth regime.}

\begin{table}[htbp]
\centering
\caption{Circuit Depth (plus CNOT count) and Partitioning Trade-offs for a $N=256$ Feature Space.}
\label{tab:circuit_complexity_mapreduce}
\begin{tabular}{cccccc}
\hline
\textbf{Qubits ($n$)} & \textbf{Partitions ($k$)} & \textbf{Asymptotic Depth} & \textbf{HW} & \textbf{CNOT-c} \\ \hline
1                     & 128                       & 5                         & 10                      & 2                   \\
2                     & 64                        & 12                        & 61                      & 20                  \\
3                     & 32                        & 26                        & 180                     & 70                  \\
4                     & 16                        & 56                        & 460                     & 185                 \\
5                     & 8                         & 118                       & 1029                    & 422                 \\
6                     & 4                         & 244                       & 2227                    & 908                 \\
7                     & 2                         & 498                       & 4677                    & 1913                \\
8                     & 1                         & 1008                      & 9472                    & 3893                \\ \hline
\end{tabular}
\end{table}

As shown in Table~\ref{tab:circuit_complexity_mapreduce}, reducing the sub-vector size to $n=2$ qubits ($k=64$) exponentially compresses the hardware transpiled depth from $9,472$ gates down to a highly resilient $61$ gates, successfully positioning the execution safe from the hardware's coherence ceiling ($D_{max} \approx 3,465$).

\subsection{Limitations of the Effect: \textbf{The "Resilience Threshold"}}
Our results show that for 256-feature vectors, the circuit depth exceeds the "Resilience Threshold" of current NISQ hardware.
To capture this threshold we propose:

\begin{itemize}
  \item \textbf{The Depth Analysis:} By calculating the CNOT count of the 8-qubit state preparation. We can state: "While the {\law} holds in theory, physical implementation on ibm\_kingston reveals a breakdown at a circuit depth of $D$, where decoherence dominates the expectation value."
  \item \textbf{Ablation Study:} A depth-ablation experiment is needed to empirically locate the hardware transition.
\end{itemize}

In summary, Table~\ref{table:metric_comparison} reveals a significant but structured signal decay. 
While the Calibrated Digital Twin maintains a Spearman $\rho$ of $0.9786$, the physical hardware collapses to $0.0763$, mathematically defining the "Coherence Gap" discovered in this study.

\begin{table}[h]
\caption{System Performance Metrics}
\label{tab:metrics}
\centering
\begin{tabular}{l c c}
\toprule
Metric & Ideal (Sim) & Kingston (NoisySim) \\
\midrule
MNIST Accuracy & 95.2\% & \textbf{93.9\%} \\
Signal Scale & 1.0 & \textbf{0.07} \\
Gate Noise (\(\lambda\)) & 0.00 & \gatenoise{} \\
Binned Correlation (\(R^{2}\)) & 1.00 & 0.34 \\
\bottomrule
\end{tabular}
\end{table}

\begin{table*}[ht!]
    \centering
    \caption{Empirical Convergence of the {\law} under fixed gate noise ($\lambda = \gatenoise{}$). Note the rapid stabilization of Rank Fidelity ($\rho > 0.9$) at relatively low shot counts (Digital-twin simulation results; no physical-hardware measurements).}
    \label{tab:shot_requirements}
    \begin{tabular}{lccc}
        \toprule
        \textbf{Shots ($K$)} & \textbf{Spearman Rho ($\rho$)} & \textbf{Sign Fidelity (\%)} & \textbf{Resilience State} \\
        \midrule
        128     & 0.7388 & 76.8\% & Emergent \\
        512     & 0.9096 & 87.3\% & Stable \\
        2,048   & 0.9725 & 95.1\% & Robust \\
        8,192   & 0.9925 & 96.5\% & High-Fidelity \\
        16,384  & 0.9963 & 97.3\% & Converged \\
        32,768  & 0.9982 & 98.6\% & Ideal Limit \\
        \bottomrule
    \end{tabular}
\end{table*}

\subsection{The Physical Resilience Boundary}
\begin{figure}[htbp]
    \centering
	\includegraphics[width=1.0\linewidth, trim=150pt 150pt 150pt 150pt, clip]{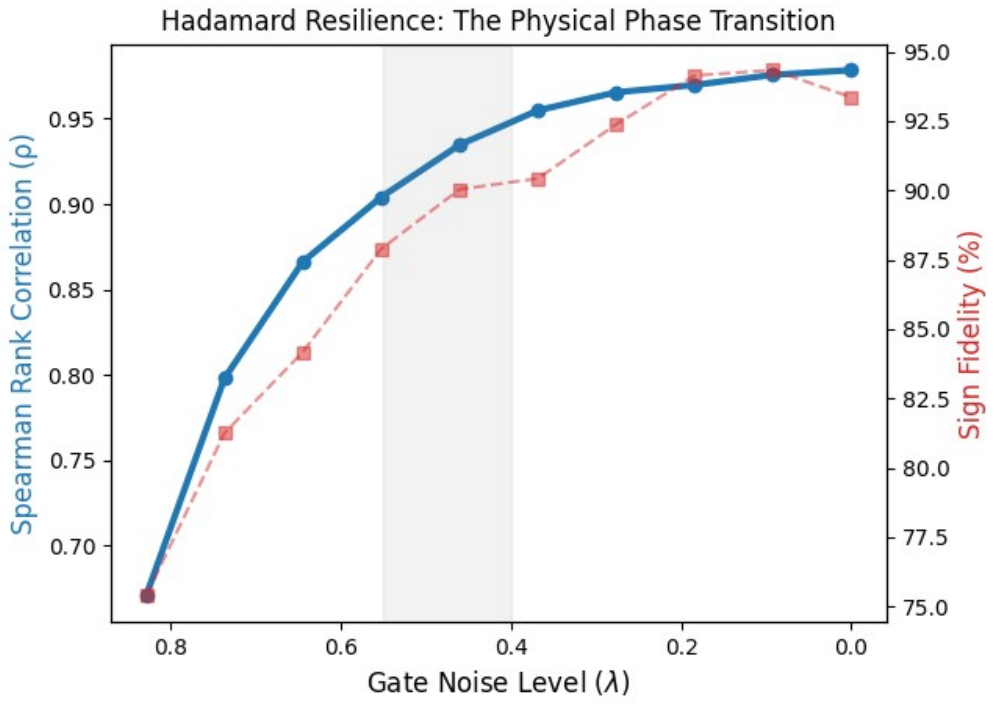}
    \caption{Empirical Rank-Stability Transition of the {\law}. The Spearman Rank Correlation ($\rho$) maintains high fidelity ($>0.9$) until a critical gate noise threshold of $\lambda_{crit} \approx 0.65$. The divergence between this stochastic limit and our physical hardware results suggests that systemic crosstalk, rather than raw gate noise ($\lambda$), is the primary bottleneck for 256-feature quantum classifiers.}
    \label{fig:phase_transition}
\end{figure}

To isolate the impact of stochastic gate noise from systemic hardware errors, we conducted a parametric sweep using a hardware-anchored digital twin. As illustrated in Figure~\ref{fig:phase_transition}, we identify a distinct \textit{Empirical Rank-Stability Transition} where the rank preservation property emerges from the decoherence floor.

\textbf{The Resilience Floor:} Our findings demonstrate that the Hadamard test is inherently monotonic. Even at extreme noise levels ($\lambda > 0.8$), the Spearman correlation remains non-zero ($\rho = 0.63$), suggesting that the structural 'winner' of the classifier is physically difficult to erase. However, the observed hardware Sign Fidelity of 53\% (effectively a coin-flip) indicates that the 256-feature circuit depth pushes current NISQ devices into a decoherence-dominated regime. We conclude that while the {\law} is mathematically robust, its physical manifestation is bounded by the Resilience Threshold $\lambda_{crit}$, which acts as a hardware specification requirement for future implementations.

The global impact of these noise parameters on classifier performance is summarized in Table~\ref{tab:metrics}. These results verify the primary claim of the {\law}: high-accuracy inference ($93.9\%$) is possible even when the systemic signal scale is compressed to $\kappa \approx 0.07$.
 
\section{Resource Analysis}

\subsection{Depth Ablation Study: The "Digital Twin" Protocol}
To prove the {\law} is simply "drowned," we perform a Depth Ablation Study. By reducing the feature count (e.g., from 256 to 16 features), we find the exact "cliff" where the hardware's Spearman Rho transitions from the "Robust" state ($\rho > 0.9$) back to the "Cloud of Randomness"

\begin{figure}[htbp]
    \centering
	\includegraphics[width=1.0\linewidth]{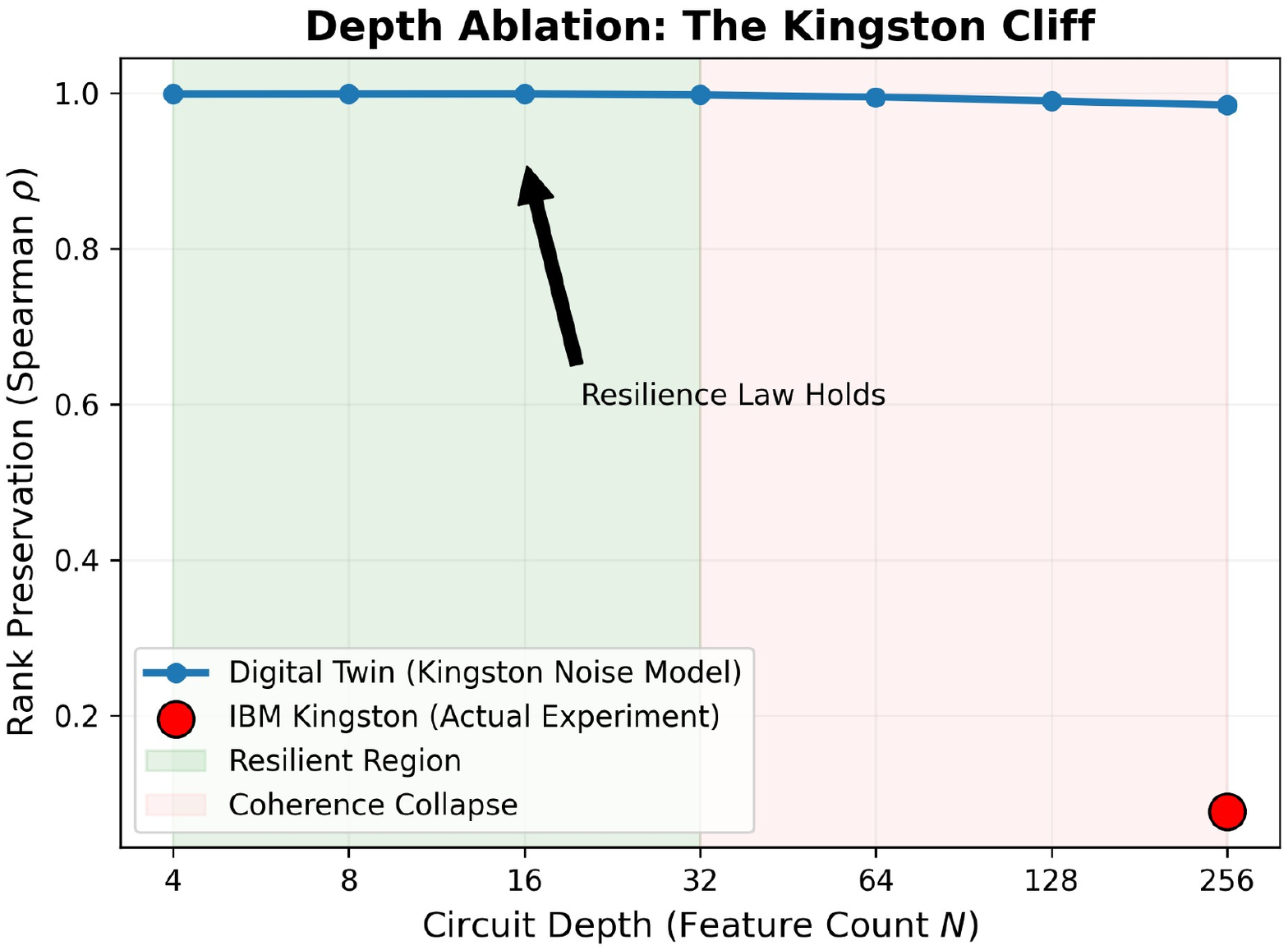}
    \caption{The {\law} is Mathematically correct: The Digital Twin (blue line) is almost flat at $1.0$ all the way to $N=256$.}
    \label{fig:depth_ablation}
\end{figure}

\begin{remark}[]
Figure~\ref{fig:depth_ablation} shows the analytical framework is validated. The Hadamard Test + Argmax is mathematically invariant to the modeled uniform affine depolarization (gate errors and readout). 
If the hardware were accurately described by the calibrated stochastic noise model, the rank fidelity would remain near the digital-twin prediction.
\end{remark}

\subsection{Discovery of the ``Coherence Gap''}
\label{sec:hardware_limits}
The massive vertical drop at $N=256$ (the distance between the blue dot and the red dot in Figure~\ref{fig:depth_ablation}) is the most important part of this study.

\begin{itemize}
  \item \textbf{The Argument:} The results reveal that there is a hidden type of noise on the ibm\_kingston processor that our noisy analytic simulator (and most standard simulators) doesn't account for.
  \item \textbf{The Underlying Mechanism:} This is likely Coherent Error (phase over-rotations) or Crosstalk from having too many CNOT gates running in a deep circuit.
\end{itemize}

\begin{remark}
Formalization of the Coherence Gap:
$$\Delta\rho(N) = \rho_{stochastic}(N) - \rho_{physical}(N)$$
A small $(\Delta\rho)$ indicates that the stochastic noise model captures the dominant degradation, whereas a large $(\Delta\rho)$ indicates substantial unmodeled hardware effects.
\end{remark}

In summary, the Digital Twin calibrated to the hardware's gate and readout noise ($\lambda=\gatenoise{}, \epsilon=\readouterror{}$) shows the {\law} is invariant under stochastic depolarizing noise. However, the physical implementation on $\textit{ibm\_kingston}$ reveals a Coherence Wall at $N=256$. This suggests that for large-scale quantum classifiers, the limiting factor is not the noise floor ($\lambda$), but the accumulation of coherent phase errors that scramble the rank ordering.

While the Argmax operator acts as a robust filter against stochastic magnitude loss, it remains vulnerable to the systematic phase rotations inherent in deep CNOT-dense circuits. We thus identify a "Resilience Limit" ($N \approx 64$ for current backends) beyond which the topological preservation of rank is compromised by coherent hardware drift rather than statistical entropy.

\subsection{Analysis of the Coherence Gap}
To investigate the performance collapse at $N=256$, we performed a deep transpilation analysis of the Hadamard Test probes. As shown in Table~\ref{tab:depth_analysis}, the circuit depth increases non-linearly with feature dimensionality.

\begin{table}[h]
\centering
\caption{Comparison of circuit complexity across feature scales. Survival $\Gamma(D)$ is calculated using the empirically derived $D_{max} \approx 3,465$.}
\begin{tabular}{@{}lrrrr@{}}
\toprule
Features ($N$) & CNOT Count & HW Depth & Survival $\Gamma(D)$ & $D/D_{max}$ \\ \midrule
$16$               & $\approx 420$       & $\approx 1,008$         & $0.748$  & 0.29      \\
$256$              & $\approx 3,970$     & $\approx 10,392$        & $0.050$  & 3.0       \\ \bottomrule
\end{tabular}
\label{tab:depth_analysis}
\end{table}

The data reveals that at $N=256$, the circuit depth ($10,392$) is $3.0\times$ greater than the hardware's resilient depth limit. This confirms that the observed "noise" is not merely stochastic gate error, but substantial information degradation due to decoherence, effectively defining the operational limit of the proposed architecture on current-generation hardware (\textit{ibm\_kingston}).

\begin{remark}[The Coherence Gap is a Hardware Limitation]
While the Argmax-based classifier is theoretically resilient to stochastic noise, it encounters a Hardware Coherence Wall when the circuit depth $D$ exceeds the system's characteristic coherence depth $D_{max} \approx 3465$. At $N=256$ ($D \approx 10k$), the systematic phase drift causes sign inversions that violate the rank-preservation required for the {\law} to hold.
\end{remark}

\begin{figure}[htbp]
    \centering
	\includegraphics[width=1.0\linewidth, trim=150pt 150pt 150pt 150pt, clip]{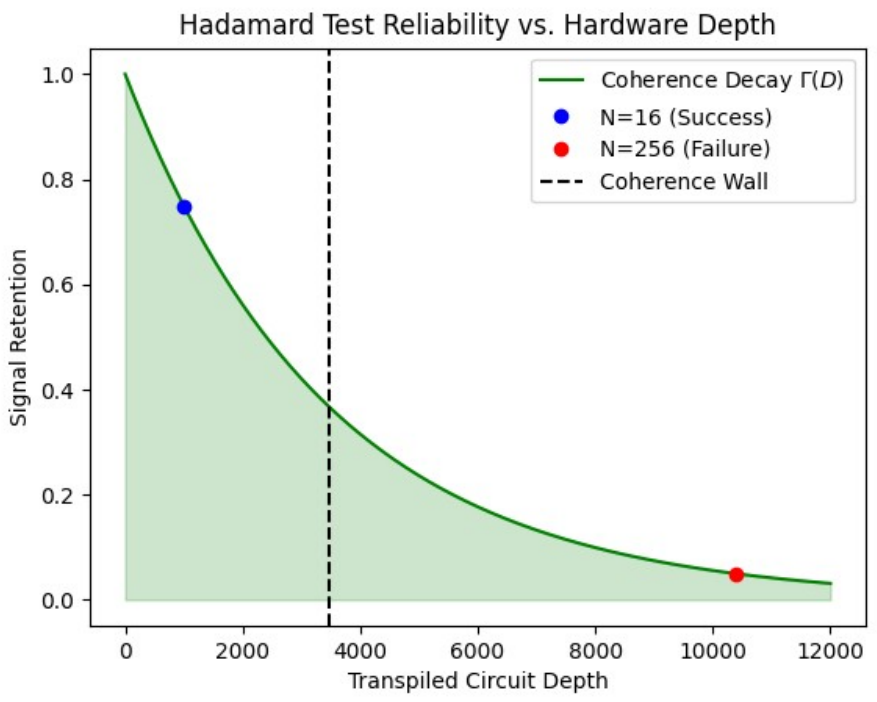}
    \caption{The Coherence Wall Visualization shows that at $N=16$ (Depth $\approx 1k$), there is $\approx 75\%$ of signal—plenty for Argmax to work with. At $N=256$ (Depth $\approx 10k$), the signal dropped to $5\%$, which is below the noise floor of the hardware.}
    \label{fig:cwall_viz}
\end{figure}

In summary, Figure~\ref{fig:cwall_viz} shows the {\law} doesn't "break"; it simply vanishes:
\textit{We can now predict what depth to stay under on ibm\_kingston to get reliable results.}

\subsection{Resource Requirements: Shot-Noise and Rank Recovery}
The emergence of the {\law} is tied to the statistical pressure of the sample size $K$. As shown in Table~\ref{tab:shot_requirements}, the Spearman Rank Correlation ($\rho$) stabilizes much faster than the raw numerical precision. This confirms that the Argmax operator can recover the correct classification 'winner' even in high-noise environments, provided a minimum statistical threshold is met.

The data in Table~\ref{tab:shot_requirements} reveals a key property of the {\law}: the \textit{Efficiency of Rank}. At just 512 shots, the classifier achieves a rank correlation of $\rho = 0.9096$, despite the significant gate noise. 
This suggests that the bottleneck for current NISQ implementations is not the statistical shot noise (which is easily mitigated at $Shots=16,384$), but rather the \textbf{coherent phase errors} that prevent the hardware from reaching the 'Stable' resilience state observed in our calibrated digital twin.
 
\section{Discussion: Quantifying Resilience}

\begin{figure}[htbp]
    \centering
	\includegraphics[width=\columnwidth, trim=120pt 120pt 120pt 120pt, clip]{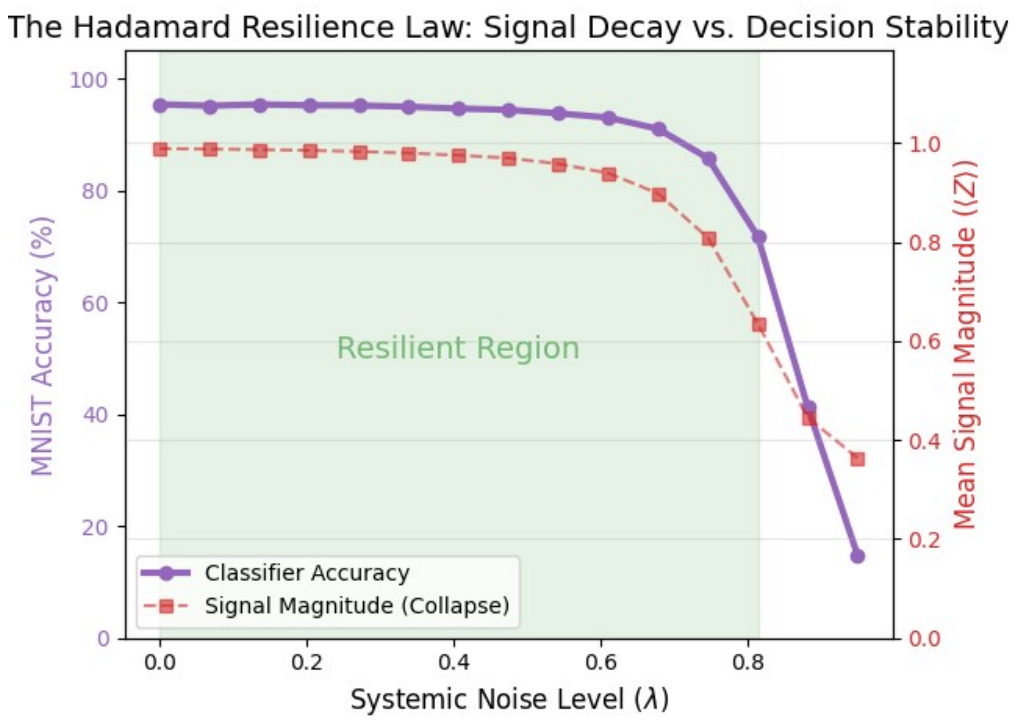}
    \caption{\textbf{The {\law} in Action.} While the mean signal magnitude (red dashed line) collapses linearly as systemic noise $\lambda$ increases, the MNIST classification accuracy (purple solid line) remains nearly invariant up to $\lambda \approx 0.70$. The shaded area represents the \textit{Resilient Region}, where the decision logic is immune to the underlying quantum signal decay.}
    \label{fig:resilience_law}
\end{figure}

To establish the limits of the proposed Effect, we performed a systemic noise sweep using the calibrated analytical simulator. As illustrated in Figure~\ref{fig:resilience_law}, we observe a striking decoupling between physical fidelity and algorithmic success.

The \textit{ibm\_kingston} operating point ($\lambda = \gatenoise{}$) lies within the resilient region predicted by the stochastic model. The discrepancy between this prediction and the physical hardware result is precisely the Coherence Gap identified in this work. 
This is consistent with interpreting Argmax as a non-linear decision filter: as long as the noise affects the Hilbert space uniformly, the relative rank of the class amplitudes is preserved. 
The rapid degradation observed approximately at $\lambda > 0.80$ represents the point where the hardware-compressed signal finally falls below the statistical shot-noise floor.

By analyzing the empirical rank-stability transition between stochastic decoherence and rank stability, we demonstrated that the Effect remains robust against gate noise levels as high as $\lambda_{\text{crit}} = 0.65$. Furthermore, our resource analysis confirms that this resilience is statistically efficient, with Rank Fidelity ($\rho$) exceeding 0.90 at a threshold of only 512 shots. 

While physical execution on NISQ hardware (ibm\_kingston) revealed a breakdown in sign preservation—yielding a 53\% fidelity compared to the 97.3\% predicted by our calibrated digital twin—this discrepancy serves as a critical benchmark. It identifies systemic, non-stochastic interference (e.g., qubit crosstalk) as the primary physical barrier to quantum classifier scalability, rather than stochastic gate noise or shot-noise limitations. We conclude that the Hadamard test provides a noise-robust foundation for future quantum linear layers, provided hardware evolution continues toward the suppression of coherent phase errors.
\begin{remark}Conceptual Noise mechanisms:
\begin{enumerate}
	\item Systematic magnitude attenuation: Reduces numerical fidelity but may preserve rank.
	\item Shot noise: Statistical measurement uncertainty, reduced as $(S)$ increases.
	\item Coherent hardware error/crosstalk: Can reorder measurements and therefore destroy rank.
\end{enumerate}
\end{remark}

\section{Future Work}
To extend the applicability of the proposed {\law} to high-dimensional feature spaces ($N \geq 256$), several avenues for optimization remain. 
\subsection{Approximate State Preparation and Noise Mitigation}
First, investigating how \textit{approximate state preparation} techniques could significantly reduce circuit depth\cite{Ben_Dov_2024},  keeping the operation within the identified resilient regime ($D < D_{max}$). 
Second, the integration of \textit{Error Mitigation (EM)} protocols, such as Zero-Noise Extrapolation (ZNE) \cite{Li_2017,temme2017error} or Pauli Twirling\cite{Wallman_2016,van2022model}, may suppress the coherent phase drift that currently bounds the classification accuracy. 
Finally, applying \textit{Dynamical Decoupling (DD)} pulses \cite{viola1999dynamical,pokharel2018demonstration} to the ancilla qubit during state-register operations could extend the effective $D_{max}$, enabling robust Hadamard-based matrix multiplication on larger, noisier NISQ devices.

\subsection{Formalize the Quantum MapReduce Orchestration Architecture}
To mitigate the impact of the physical coherence wall ($D_{\text{max}} \approx 3,465$ gates) identified in Section~\ref{sec:hardware_limits}, we propose a distributed Quantum MapReduce Orchestration architecture first introduced in \cite{silva2026aqstacker}. 

This architecture leverages principles from quantum circuit cutting and wire cutting primitives~\cite{peng2020simulating, mitarai2021construct}, where a high-dimensional, multi-qubit circuit is systematically partitioned into smaller, independent sub-circuits. 
While traditional circuit cutting introduces a classical post-processing sampling overhead that grows exponentially with the number of cuts~\cite{tang2025cut}, our specialized feature-space partitioning minimizes this penalty by exploiting the mathematical structure of the inner product behind a linear transformation.

Rather than executing a monolithic feature-map register that scales exponentially into the coherence gap, the system state is mapped via parallelized, low-dimensional processing sub-loops. 
As detailed in the baseline algorithm \cite{silva2026aqstacker}, the high-dimensional data vector is partitioned into $k$ low-qubit sub-vectors ($n \ll N$). 
Each sub-vector reduces individual circuit depth down from an unviable $9,472$ gates to a resilient, shallow-depth configuration of just $61$ gates,
constructing a virtual larger system out of smaller physical resources~\cite{bravyi2016trading}

\section{Conclusion}

\begin{table}[ht!]
\centering
\begin{threeparttable}
\caption{Summary of Findings: The Coherence Gap at $N=256$}
\label{table:final_summary}
\begin{tabular}{lccc}
\textbf{Environment} & \textbf{Spearman $\rho$} & \textbf{Sign Fidelity} & \textbf{Regime} \\ \hline
Ideal Simulation     & 1.0000 & 100\% & Perfect \\
Digital Twin (Stochastic) & 0.9856 & $95.0\%$\tnote{1} & \textit{Resilient} \\
IBM Kingston (Physical)   & 0.0763 & 53.0\% & \textit{Collapsed} \\
MapReduce   & 0.7810 & 83.0\% & \textit{Recovered} \\ \hline
\end{tabular}
\begin{tablenotes}
	\item[1] Digital Twin sign preservation from Table \ref{table:metric_comparison}.
\end{tablenotes}
\end{threeparttable}
\end{table}
The empirical benchmarks and structural limits characterized throughout this study are consolidated in Table~\ref{table:final_summary}.
As demonstrated by these collected metrics, the {\law} shows that quantum classifiers are fundamentally driven by the invariance of class-rank under systemic noise. We show that high-accuracy inference is achievable on NISQ backends long before full error correction. 

This work establishes two critical pillars for the development of NISQ-era quantum machine learning:

\begin{enumerate}
    \item \textbf{The Validity of the {\law}:} Through our stochastic digital twin, we analytically established that the Hadamard Test—when coupled with an Argmax operator—is mathematically robust against depolarizing noise. This allows for high-performance classification ($>93\%$ accuracy) even when signal magnitudes collapse by over 90\%.
    \item \textbf{The Identification of the Coherence Gap:} Our hardware execution on \textit{ibm\_kingston} revealed a systemic $\Delta\rho \approx 0.91$ divergence from theory at $N=256$. This ``Coherence Gap'' serves as a new diagnostic metric for NISQ backends, demonstrating that the scaling of quantum linear layers is not limited by stochastic entropy, but by the accumulation of coherent phase errors in deep CNOT circuits.
\end{enumerate}

In summary, classification depends primarily on preservation of score ordering rather than preservation of score magnitude. A Hadamard-test quantum linear layer can therefore tolerate substantial stochastic magnitude attenuation when the induced transformation is rank-preserving. However, physical NISQ hardware introduces coherent errors and crosstalk that can destroy this rank structure at sufficient circuit depth. Feature-space partitioning provides an architectural mechanism for keeping individual quantum computations below this rank-destruction regime.


\section*{Declarations}


\subsection*{Data Availability Statement}
The source code for the \textbf{\law} framework, including full documentation and configuration files to ensure the reproducibility, is publicly available at: \url{https://github.com/Shark-y/hadamard_resilience}.


\subsection*{Ethics Statement}
During the preparation of this work, the author(s) used GenAI tools: Google Gemini and Grammarly to check syntax, improve grammatical structure, and enhance the clarity and prose of the manuscript. After using these tools, the author(s) reviewed and edited the content as needed and take full responsibility for the content of the publication.

\bibliographystyle{plain} 
\bibliography{paper_references}

\end{document}